\documentclass[12pt,letterpaper]{article}

\parskip=\medskipamount            
\def\7#1#2{\mathop{\null#2}\limits^{#1}}        

\def\beee{\begin{equation}}
\def\eeee{\end{equation}}

\begin{document}

\bibliographystyle{unsrt}
\begin{center}
\textbf{COVARIANCE OF TIME-ORDERED PRODUCTS\\
IMPLIES LOCAL COMMUTATIVITY OF FIELDS}\\
[5mm]
O.W. Greenberg\footnote{email address, owgreen@physics.umd.edu.}\\
{\it Center for Theoretical Physics\\
Department of Physics \\
University of Maryland\\
College Park, MD~~20742-4111}\\
~\\

\end{center}

\begin{abstract}

We formulate Lorentz covariance of a quantum field theory in terms of
covariance of time-ordered products (or other Green's functions). This
formulation of Lorentz covariance implies
spacelike local commutativity or anticommutativity of fields,
sometimes called microscopic
causality or microcausality. With this formulation microcausality
does not have to be taken as a separate assumption.

\end{abstract}

\section{Introduction}

In formulating the basic assumptions of relativistic quantum
field theory in terms of vacuum matrix elements of products of
fields, A.S. Wightman~\cite{wig,wig2} chose to use unordered products
of fields and to express the basic assumptions in terms of the
Wightman distributions or, colloquially, functions. We will use
commutativity (commutator) both for commutativity (commutator)
for Bose fields  and anticommutativity (anticommutator) for Fermi
fields throughout this paper. In Wightman's formulation spacelike
commutativity of fields is independent of relativistic covariance
of the Wightman functions. It is generally accepted that local
commutativity implies that vacuum matrix elements of
time-ordered products of fields (or other Green's functions) are
Lorentz covariant. The converse, that Lorentz covariance of
vacuum matrix elements of time-ordered products of fields implies
local commutativity of fields, does not seem to appear in the
field theory literature. If we take Lorentz covariance of
time-ordered products as the condition of Lorentz covariance of
the field theory, then the demonstration of this converse, which
we give below, means that local commutativity is not an
independent assumption of the theory. This implies further that
the spin-statistics and $\mathcal{CPT}$ theorems also hold
without further assumptions. All of this means that relativistic
quantum field theory becomes a more coherent structure.

\section{Proof that covariance of time-ordered products implies
local commutativity of fields}

Consider an arbitrary time-ordered function
\begin{eqnarray}
\lefteqn{\tau(x_{-N_L}, \cdots, x_{-1}, x_{-0}, x_{0}, x_{1},
\cdots, x_{N_R})  = }     \nonumber  \\
& & \langle 0|T(\phi(x_{-N_L}) \cdots \phi(x_{-1}) \phi(x_{-0})
\phi(x_{0}) \phi(x_1) \cdots \phi(x_{N_R}))|0 \rangle.     \label{t}
\end{eqnarray}
There are $N_L + N_R + 2$ points. Choose these points so that they
correspond to a Jost point~\cite{jos}, i.e.,
the successive difference variables
\begin{eqnarray}
\lefteqn{
\xi_{-N_L}=x_{-N_L}-x_{-N_L+1},
\cdots, \xi_{-1}=x_{-1}-x_{-0}, }   \nonumber   \\
& & ~\xi=x_{-0}-x_{0},~\xi_{1}=x_{0}-x_{1},
\cdots, ~\xi_{N_R}=x_{N_R-1}- x_{N_R}
\end{eqnarray}
form a convex set that is totally spacelike.\footnote{My metric is 
$\mathrm{diag} (1, -1,-1,-1)$
and I use $x^2=(x^0)^2-(x^1)^2-(x^2)^2-(x^3)^2$ as short for $x \cdot x$. I
abbreviate $x^2 < 0$, i.e. spacelike, by $x \sim 0$; $x^2 >0, x^0 >0$, i.e.,
the open positive light cone, by $x \in V_+$; the open negative light cone by
$x \in V_-$; and the closed light cones by $\bar{V_{\pm}}$.}
That requires
\beee
\sum_{l=-1}^{-N_L} \lambda_l \xi_l + \lambda \xi +
\sum_{r=1}^{N_R} \lambda_r \xi_r \sim 0, \forall \lambda_{l} \geq 0,
\forall \lambda_{r} \geq 0, ~ \lambda \geq 0,
\eeee
and
\beee
\sum_{l=1}^{N_L-1} \lambda_{l} +\lambda + \sum_{r=1}^{N_R-1}  \lambda_{r} >0.
\eeee
Because we chose the points to correspond to a Jost point, we can equate the
time-ordered function of Eq.(\ref{t}) to a Wightman function, i.e. to the vacuum
matrix element of a product of fields,
\begin{eqnarray}
\lefteqn{\mathcal{W}(x_{-N_L},\cdots, x_{-1},x_{-0},x_0,x_1, \cdots, x_{N_R})} \nonumber \\
& & =\langle 0| \phi(x_{-N_L}) \cdots  \phi(x_{-1}) \phi(x_{-0}) 
\phi(x_0) \phi(x_1)\cdots 
\phi(x_{N_R}) |0\rangle.     \label{w}
\end{eqnarray}
Choose the points $x_i$ such that under a Lorentz transformation
the time order of $x_{-0}$ and
$x_{0}$ will reverse without changing the relative time order of any of the other
points. Observer Lorentz covariance of the time-ordered product requires 
that these two different
time orders must agree; this means the commutator
must vanish at spacelike separation at Jost points in each Wightman function.
These points will be sufficiently separated so that the same condition
will hold for an open neighborhood of these points. Now add imaginary parts,
$\eta_i \in V_-$, i.e. in or on the backward light cone in momentum space,
to all the difference variables except $\xi$ to define complex difference
variables
\begin{eqnarray}
\lefteqn{
\zeta_{-N_L}=\xi_{-N_L}+i \eta_{-N_L}, \cdots,
\zeta_{-1} = \xi_{-1}+i \eta_{-1};   } \nonumber \\
& &\zeta_{1} =\xi_{1}+i \eta_{1},
\cdots, \zeta_{N_R}=\xi_{N_R}+i \eta_{N_R}.
\end{eqnarray}
Note that we do not choose a complex variable to correspond to
$\xi=x_{-0}-x_{0}$. The complex parts of the difference variables are
associated with the vectors $x_{-l}$ and $x_{r}$ and not with the
vectors $x_{-0}$ and $x_{0}$.
Then this condition, that
\beee
\langle 0|\phi(x_{-N_L}) \cdots \phi(x_{-1}) [\phi(x_{-0}), \phi(x_0)]_{\mp}
\phi(x_{1})n\cdots \phi(x_{N_R})|0 \rangle = 0,
\eeee
can be analytically continued
to all $\zeta_{-l}$ and $\zeta_{r}$. The boundary values for
$\mathrm{Im}\zeta_{-l} \rightarrow 0$ and $\mathrm{Im}\zeta_{r} \rightarrow 0$
are then equal for all $x_{-l}$ and $x_r$ and for $x_{-0}$ spacelike with respect
to $x_0$; i.e. an arbitrary matrix element of the commutator
$[\phi(x_{-0}),\phi(x_0)]_{\mp}=0$ for
spacelike separation. Wightman's reconstruction theorem~\cite{wig3,jos2} states that the set of all
Wightman functions determines the quantum field theory up to unitary 
equivalence. We do not repeat the conditions for the reconstruction theorem 
here; they are the usual conditions of a relativistic quantum field theory 
and are described in detail in~\cite{wig,wig2,jos2}. Thus the field is local and the theory
obeys microcausality. Similar arguments show that the same conclusion
follows from covariance of other Greens functions.\footnote{Note that if 
we had tried to make the same argument using an arbitrary time-ordered
function we would have been able to conclude that 
\begin{eqnarray}
\lefteqn{\langle 0|T(\phi(x_{-N_L}) \cdots
\phi(x_{-1}) \phi(x_{-0}) \phi(x_{0})\phi(x_{1}) 
\cdots \phi(x_{N_R}))|0 \rangle  =}    \nonumber \\
& &  \langle 0|T(\phi(x_{-N_L}) \cdots \phi(x_{-1}) \phi(x_{0}) \phi(x_{-0})
\phi(x_{1}) \cdots \phi(x_{N_R}))|0 \rangle;  \label{tt}   
\end{eqnarray}
however because the reconstruction theorem holds for Wightman functions, but not
for time-ordered functions, we would not be able to conclude that we could 
reconstruct the quantum field theory. The merit of the argument using the 
time-ordered product at Jost points is that it allows the transition to 
Wightman functions. The analyticity of the Wightman functions in the neighborhood
of a Jost function allows the conclusion that arbitrary matrix elements of the 
commutator (or anticommutator) vanish at spacelike separation and, using the
reconstruction theorem, that the arbitrary matrix elements define a local quantum
field theory. It may be that there is an analog of the reconstruction theorem for
time-ordered products, but because the
step functions that occur in the time-ordered product are 
discontinuous, the definition of a time-ordered product in terms of a Wightman
function has ambiguities, and one can expect that such an analog will have technical
complications.}

Choose the $N_L + N_R + 2$ points as follows,
\begin{eqnarray}
\lefteqn{x_l=(-la,0,0,-3la), 1 \leq l \leq N_L};   \nonumber   \\
& &
x_{r}=(ra,0,0,3ra),
~1 \leq r \leq N_R,
 x_{-0}= (0, 0, 0, -a), ~x_0 = (0, 0, 0, a).
\end{eqnarray}
We can check that these points are Jost points and that there exists an open
set containing them that contains only Jost points. Further, a small boost in the
$\pm z$-direction will make $x_0^0$ either greater than or less than
$x_{-0}^0$ without changing the time ordering of any of the other points.
In order for the time-ordered function (distribution) to be independent of
the frame of reference of the observer, the commutator must
vanish at spacelike separation of $x_{-0}$ and $x_0$.

Our analysis uses properties of the time-ordered product and the
commutator at spacelike separation only and therefore does not depend
on the difference between the time-ordered product and the
$T^{\star}$-product which occurs only at coincident points~\cite{jac,gro,das}.

We emphasize that our demonstration does not assume pointlike form of the
Lagrangian or Hamiltonian, therefore our argument holds for nonlocal theories
in which the fields in the Lagrangian or Hamiltonian enter at separated
points in spacetime.

These arguments do not hold for parastatistics or quon fields because the
time-ordered products for those fields are not simply related when
neighboring fields are
interchanged at spacelike separation. Observables in theories with
parastatistics fields, but not in theories with quon fields, commute at
spacelike separation, but the parastatistics fields themselves do not 
commute at spacelike separation. 

The discussion given above for bose and fermi fields can be carried over
word for word to conclude that observer covariance of time-ordered products
of observable fields implies local commutativity (for observable fields
always commutativity and not anticommutativity) of observable fields. This
argument applies to observable fields in theories with
bose, fermi, parabose and parafermi fields.

Acknowledgements: I am happy to thank Lev Okun, Peter Orland
and Ching-Hung Woo for valuable discussions.
This work was supported in part by the National Science Foundation,
Award No. PHY-0140301.

\end{document}